\documentclass[%
aps,preprint,prc,amsmath,amssymb,nofootinbib,superscriptaddress%
]{revtex4}

\usepackage[utf8]{inputenc}
\usepackage[usenames]{color}
\usepackage[dvipsnames]{xcolor}
\usepackage{graphicx}
\usepackage{amsmath}
\usepackage{amsfonts}
\usepackage{amssymb}
\usepackage{mathrsfs}
\usepackage{bm}
\usepackage{verbatim}
\usepackage{cancel}
\usepackage{comment}
\usepackage[normalem]{ulem}
\usepackage{float}
\usepackage{textcomp}
\usepackage{booktabs}
\usepackage{braket}

\usepackage{hyperref}
\hypersetup{
  colorlinks=true, 
  linkcolor=ForestGreen,
}

\usepackage{multirow,makecell}

\newcommand{\chn}[3]{{{}^{#1}\!{#2}_{#3}}}
\newcommand{\cs}[2]{\chn{#1}{S}{#2}}
\newcommand{\chp}[2]{\chn{#1}{P}{#2}}
\newcommand{\cd}[2]{\chn{#1}{D}{#2}}
\newcommand{\cf}[2]{\chn{#1}{F}{#2}}

\newcommand{\csd}{{\cs{3}{1}-\cd{3}{1}}}
\newcommand{\cpf}{{\chp{3}{2}-\cf{3}{2}}}

\newcommand{\VSOI}[1]{{\widehat{\mathcal{O}}_{#1}}}
\newcommand{\NNLO}{N$^2$LO}

\begin{document}

\title{Perturbative renormalization of chiral nuclear forces at subleading order in $\csd$ channel}

\author{Rui Peng}
\affiliation{School of Physics, and State Key Laboratory of Nuclear Physics and Technology, Peking University, Beijing 100871, China}

\author{Bingwei Long}
\email{bingwei@scu.edu}
\affiliation{College of Physics, Sichuan University, Chengdu 610065, China}
\affiliation{Southern Center for Nuclear-Science Theory (SCNT), Institute of Modern Physics, Chinese Academy of Sciences, Huizhou 516000, Guangdong Province, China}

\author{Fu-Rong Xu}
\email{frxu@pku.edu.cn}
\affiliation{School of Physics, and State Key Laboratory of Nuclear Physics and Technology, Peking University, Beijing 100871, China}
\affiliation{Southern Center for Nuclear-Science Theory (SCNT), Institute of Modern Physics, Chinese Academy of Sciences, Huizhou 516000, Guangdong Province, China}

\date{August 7, 2025}

\begin{abstract}

We investigate renormalization of chiral nuclear forces in the coupled channel of $\csd$ of nucleon-nucleon scattering.
The one-pion exchange potential is treated nonperturbatively at leading order while subleading potentials are perturbations. 
Very much like the uncoupled channel of $\chp{3}{0}$, the singular attraction of one-pion exchange gives rise to the so-called genuine exceptional cutoffs, where artificial correlations between subleading contact operators emerge and they result in ill-defined values of the low-energy constants.
To address this issue we follow the solution proposed for $\chp{3}{0}$ in Ref.~\cite{Peng:2024aiz} and apply it to $\csd$.
The truncation uncertainty of an effective field theory allows certain degrees of freedom in choosing renormalization conditions, or fitting schemes of the low-energy constants.
By exploiting this freedom near the exceptional cutoffs, we are able to remove the said correlations. 
A much mitigated cutoff variation of the phase shifts, which is acceptable to the power counting, is thus obtained.

\end{abstract}
\maketitle

\section{Introduction\label{sec:intro}}

Within the framework of chiral effective field theory (ChEFT), renormalization group (RG) invariance is widely considered an important guideline for constructing nucleon-nucleon ($NN$) forces~\cite{Kaplan:1996xu, Nogga:2005hy, Birse:2005um, Long:2011xw, Long:2012ve, PavonValderrama:2016lqn, Bira:2020rv} and nuclear currents~\cite{PavonValderrama:2014zeq, Shi:2022blm, Liu:2022cfd} and studying fundamental symmetries~\cite{deVries:2020loy, Cirigliano:2018hja, Cirigliano:2020dmx, zhaopw:ndbd}. 
Renormalization-group invariance requires that the observables be independent of the arbitrarily chosen momentum cutoff $\Lambda$ introduced to regularize the ultraviolet behavior of the nuclear interactions.
Renormalization is the procedure in which a sufficient number of contact potentials (or counterterms) are present and their low-energy constants (LECs) run with $\Lambda$ so as to absorb, if any, the $\Lambda$ dependence of the observables.
One-pion exchange (OPE), the dominant part of long-range chiral nuclear forces, poses challenges to renormalization due to its singular and attractive tensor force in spin-triplet channels, such as $\csd$, $\chp{3}{0}$ and $\cpf$, among others. 
For instance, the singular part of the OPE potential ($\sim -1/r^3$) causes the $NN$ system to collapse~\cite{Beane:2000wh, Beane:2001bc, Nogga:2005hy}.
This collapse also presents itself as an extreme sensitivity to  $\Lambda$ of the phase shifts.
Therefore, if OPE is treated nonperturbatively at leading order (LO), a short-range counterterm must be placed concurrently at LO in each of these channels to ensure the stability of the $NN$ system and to preserve RG invariance.

In the present paper, we focus on renormalization in the coupled channel of $\csd$ at the subleading order, where the subleading amplitude is calculated by the distorted-wave expansion.
The same problem was originally tackled in Refs.~\cite{Valderrama:2009ei, PavonValderrama:2011fcz, Long:2011xw} in which a set of subleading counterterms are identified under the guidance of RG invariance.
But in light of a recent study of singular attractive potentials~\cite{Gasparyan:2022isg}, renormalization of $\csd$ needs a revisit.
In Ref.~\cite{Gasparyan:2022isg}, a detailed scan of $\Lambda$ in $\chp{3}{0}$ discovers the so-called genuine exceptional cutoffs (GECs) near which the subleading amplitude appears to be diverging, thus violating RG invariance.

We have addressed this problem for uncoupled channels in a previous work~\cite{Peng:2024aiz}, using $\chp{3}{0}$ as the example. 
With the potential at next-to-leading order (NLO) vanishing, the first nontrivial correction to the $\chp{3}{0}$ phase shifts starts at next-to-next-to-leading order ({\NNLO}) and it is generated by a linear combination of the contact potentials $\mathcal{O}_0$ and $\mathcal{O}_2$, in addition to the two-pion exchange (TPE) potential.
The LECs of $\mathcal{O}^{(0)}$ and $\mathcal{O}^{(2)}$--- $C^{(2)}$ and $D^{(0)}$--- are determined by fitting to the phase shifts at the center-of-mass momentum $k_1$ and $k_2$.
It is found there that at a GEC $\Lambda_E$ the phase-shift corrections generated by $\mathcal{O}^{(0)}$ and $\mathcal{O}^{(2)}$, which are otherwise independent operators, become correlated. 
This correlation between $k_1$ and $k_2$ makes it impossible to determine $C^{(2)}$ and $D^{(0)}$ from the phase shifts at $k_1$ and $k_2$ because they do not provide sufficient ``signals'' to tell apart the two LECs.
This correlation appears to be an artifact produced by the ultraviolet regulator because the value of $\Lambda_E$, for a given pair of $k_1$ and $k_2$, are quite sensitive to the choice of regulators.
Indeed, Ref.~\cite{Yang:2024yqv} proposes to change the regulator around $\Lambda_E$ which effectively breaks this correlation. 
In addition, the possibility of finding a class of regulators that do no at all have any GEC is not ruled out~\cite{PavonValderrama:2025azr}. 

Although the issue reported in Ref.~\cite{Gasparyan:2022isg} is not a conceptual problem, it poses a challenge to the determination of subleading LECs.
We propose in Ref.~\cite{Peng:2024aiz} to resolve it by adjusting the strategy of extracting the LECs near $\Lambda_E$, thereby restoring RG invariance. 
The modified fitting strategy makes use of the inherently approximate nature of EFT, removing the correlation by a small tuning of the LO LEC. 

The coupled-channel dynamics complicates the correlation issue significantly because now we have to examine three scattering parameters at a single value of $k$: two phase shifts and one mixing angle.
Fortunately, $\csd$ is the only coupled channel one needs to study carefully. 
As shown in Ref.~\cite{Wu:2018lai} that the OPE potential is suppressed so much by the centrifugal barrier in $\cpf$ and higher partial waves that it is quite perturbative; therefore, the complex renormalization problem due to nonperturbative OPE is no longer a concern in those channels.
Although the solution we offer in the end is more complicated than that of $\chp{3}{0}$, the methodology is similar.
We want to be able to extract the values of subleading LECs by making small adjustments to how the empirical phase shifts (and other observables) are taken as the inputs.

This paper is organized as follows. 
In Sec.~\ref{sec:LO}, we outline how to calculate the LO $\csd$ amplitude and the deuteron binding energy.
The perturbative treatment of the subleading interactions is explained in Sec.~\ref{sec:NNLO} and how the GECs arise in determination of the LECs is detailed in Sec.~\ref{sec:exceptional}.
We then discuss in Sec.~\ref{sec:solution} how to resolve this issue by modifying the fitting strategy near the GECs, which is followed by a summary in Sec.~\ref{sec:sum}.

\section{Leading order \label{sec:LO}}

The LO amplitude is obtained by the full iteration of the LO potential $V^{(0)}$, via solving the partial-wave Lippmann-Schwinger (LS) equation exactly. In the coupled channel of $\csd$, the scattering amplitude is a $2 \times 2$ matrix, and its elements are given by
\begin{equation}
    \begin{aligned}
          T^{(0)}_{L'L}(p',p;k) &\equiv \bra{\csd,p';k,L'}T^{(0)}\ket{\csd,p;k,L} \, \\
        &= V^{(0)}_{L'L}(p',p)+\sum_{L''=0,2}\int dl\, l^2V^{(0)}_{L'L''}(p',l)\frac{T_{L''L}^{(0)}(l,p;k)}{k^2-l^2+i\epsilon} \, ,
    \end{aligned}
    \label{equ:LSE}
\end{equation}
where $L$ ($L'$) and $p$ ($p'$) are the initial (final) orbital angular momentum and relative momentum, and $k$ the on-shell center-of-mass momentum. 
In this work, the following values of physical constants are used: the nucleon mass $m_N = 938$ MeV, pion mass $m_\pi=138$ MeV, pion decay constant $f_\pi = 92.4$ MeV and the nucleon axial coupling $g_A = 1.29$.
The potentials are regularized by a separable, ultraviolet regulator:
\begin{equation}
    V(p',p; \Lambda) \equiv f_R(\frac{p'^2}{\Lambda^2})V(p',p)f_R(\frac{p^2}{\Lambda^2}) \, ,
    \label{equ:G_regulator}
\end{equation}
where $f_R$ is the Gaussian regulator, given by 
\begin{equation}
    f_R(x) = e^{-x^2} \, .
\end{equation}

$V^{(0)}$ has two parts: one is the long-range force of OPE $V_\pi$ and the other is the short-range, or contact, potential $V_S^{(0)}$. 
The partial-wave decomposition of long-range potentials will be useful and can be found in Refs.~\cite{Kaiser:1997mw, Epelbaum:1999dj}.
For the short-range potential, we follow Refs.~\cite{Bira:1996, Kaplan:1996xu, EPELBAOUM1999413, Long:2012ve, Long:2011xw, Long:2011qx, Peng_2020} to write all the contact potentials in the partial-wave basis.
The LO contact potential in $\csd$ is given by
\begin{equation}
    V_S^{(0)} =  C^{(0)} \VSOI{C} \, ,
\end{equation}
where $C^{(0)}$ is the LO LEC and $\VSOI{C}$ is defined as
\begin{equation}
    \bra{\csd, p'} \VSOI{C} \ket{\csd, p} =\begin{pmatrix}
        1  &  0 \\
        0  &  0 \\
    \end{pmatrix} \, . 
    \label{eqn:LOVS}
\end{equation}

The LS equation~\eqref{equ:LSE} is commonly expressed in the following schematic form:
\begin{equation}
    T^{(0)} = V^{(0)} + V^{(0)} G_0 T^{(0)} \, .
\end{equation}
The LO $S$-matrix is related to the LO on-shell $T$-matrix by
\begin{equation}
    S^{(0)} = 1 - i\pi kT^{(0)} \, .
\end{equation}
The $\csd$ phase shifts are related to the $S$-matrix by the Stapp parameterization~\cite{Stapp1957}:
\begin{equation}
    S^{(0)} = \begin{pmatrix}
        \mathrm{cos}(2\varepsilon^{(0)})e^{2i\delta_1^{(0)}} & i\mathrm{sin}(2\varepsilon^{(0)})e^{i\delta_1^{(0)}+\delta_2^{(0)}} \\
        i\mathrm{sin}(2\varepsilon^{(0)})e^{i\delta_1^{(0)}+\delta_2^{(0)}} & \mathrm{cos}(2\varepsilon^{(0)})e^{2i\delta_2^{(0)}}  
    \end{pmatrix} \, ,
\end{equation}
where $\delta_1^{(0)}$ and $\delta_2^{(0)}$ represent the LO phase shifts of $^3S_1$ and $^3D_1$, respectively, and $\varepsilon^{(0)}$ the LO mixing angle.

We determine the value of $C^{(0)}$ by the deuteron binding energy, the bound state in $\csd$.
The binding energy is derived from the homogeneous version of Eq.~\eqref{equ:LSE}:
\begin{equation}
    \psi_L^{(0)}(p) = \frac{1}{-m_NB^{(0)}-p^2} \sum_{L'=0,2} \int dp'\, p'^2V_{LL'}^{(0)}(p,p') \psi_{L'}^{(0)}(p') \, ,
    \label{equ:BDELO}
\end{equation}
where $\psi_L^{(0)}$ is the LO bound state wave function and $B^{(0)}$ the LO binding energy. Except for a small cutoff window near the GECs,  as will be explained later, we fit $B^{(0)}$ to the empirical deuteron binding energy $B_\text{emp} = 2.225$ MeV.

Whether fitting to the binding energy or the phase shift at a chosen energy, $C^{(0)}(\Lambda)$ exhibits a limit-cycle-like behavior as shown in Ref.~\cite{Nogga:2005hy}. It diverges at some cutoff values, which are denoted by $\Lambda_\star$ in the paper.
We note that below $3$ GeV there are two diverging points of $C^{(0)}(\Lambda)$: $\Lambda_\star = $ 1032 MeV and 2953 MeV. 
These divergences are correlated with the emergence of spurious deeply bound states.
The running of a coupling constant is not observable and may differ much from one regulator to another.
In fact, Ref.~\cite{Beane:2000wh} shows, as also noted in Ref.~\cite{Nogga:2005hy}, that with a local, coordinate-space regulator $C^{(0)}(\Lambda)$ does not really diverge.

\section{Subleading potentials as perturbations\label{sec:NNLO}}

We follow the power counting established in Ref.~\cite{Long:2011xw} where the NLO is argued to vanish in $\csd$ channel and the first nontrivial correction to LO appears at {\NNLO} ($\mathcal{O}(Q^2)$). 
The short-range potential at {\NNLO}, $V_S^{(2)}$, is the linear combination of the following counterterms:
\begin{equation}
    V_S^{(2)} =  C^{(2)} \VSOI{C} + D^{(0)} \VSOI{D} + E^{(0)} \VSOI{E} \, ,
\end{equation}
where $C^{(2)}$, $D^{(0)}$ and $E^{(0)}$ are the LECs at {\NNLO}, and $\VSOI{D}$ and $\VSOI{E}$ are given by
\begin{equation}
    \begin{aligned}
        \bra{\csd, p'} \VSOI{D} \ket{\csd, p} =& \begin{pmatrix}
        p^2+p'^2  &  0 \\
        0  &  0 \\
    \end{pmatrix} \, , \\
    \bra{\csd, p'} \VSOI{E} \ket{\csd, p} =& \begin{pmatrix}
        0  &  p^2 \\
        p'^2  &  0 \\
    \end{pmatrix} \, .
    \end{aligned}
\end{equation}

The long-range part of the {\NNLO} potential is the TPE potential composed of vertexes with chiral index $\nu = 0$, $V_{2\pi}^{(0)}$. For clarity, we list the $\csd$ potentials order by order up to {\NNLO} as follows:
\begin{equation}
\begin{aligned}
    V^{(0)} &= V_\pi + V_S^{(0)} \, , \\
    V^{(1)} &= 0 \, , \\
    V^{(2)} &= V_{2\pi}^{(0)} + V_S^{(2)} \, .
\end{aligned}
\end{equation}
As previously stated, $V^{(0)}$ is treated nonperturbatively by solving the LS equation whereas $V^{(2)}$ is treated perturbatively on top of the LO amplitude. Therefore, the {\NNLO} correction to the amplitude, $T^{(2)}$, is given by
\begin{equation}
    T^{(2)} = (1 + T^{(0)} G_0) V^{(2)} (1 + G_0 T^{(0)}) \, ,
    \label{equ:T2_pert}
\end{equation}
which can also be expressed as the first order in the distorted-wave expansion: 
\begin{equation}
    T^{(2)}(k) = \bra{\psi_k^-, \csd} V^{(2)} \ket{\psi_k^+, \csd} \, .
    \label{equ:T2_psi}
\end{equation}
Here, $\psi_k^+$ ($\psi_k^-$) denotes the in-state (out-state) scattering states.
Using the distorted-wave expansion of the $S$-matrix, as outlined in the appendix of Ref.~\cite{Long:2011xw}, we can relate the {\NNLO} $^3S_1$ phase shift correction $\delta_1^{(2)}$ and the mixing angle correction $\varepsilon^{(2)}$ to $T^{(2)}$ as follows:
\begin{equation}
    \begin{aligned}
        \delta_1^{(2)}(k) &= \frac{-\pi k}{2\mathrm{cos}(2\varepsilon^{(0)})}Re\left[e^{-2i\delta_1^{(0)}}T_{00}^{(2)}(k)\right] \, , \\
        \varepsilon^{(2)}(k) &= \frac{-\pi k}{2\mathrm{cos}(2\varepsilon^{(0)})}Re\left[e^{-i(\delta_1^{(0)}+\delta_2^{(0)})}T_{02}^{(2)}(k)\right] \, , \\
    \end{aligned}
\end{equation}
where the subscripts of $T^{(2)}_{L' L}$ indicate the orbital angular momenta of the initial and final states.

Equation~\eqref{equ:T2_psi} suggests that $T^{(2)}$ is a linear combination of the contributions from the operators of $V^{(2)}$. 
As a result, the LECs appearing as the coupling constants of these operators can be extracted by a linear system of equations that equates the {\NNLO} EFT phase shifts with their empirical values. 
Introducing the following quantities,
\begin{equation}
    \begin{aligned}
        \mathcal{T}^{C,D,E}(k) & \equiv \bra{\psi_k^-, \csd} \VSOI{C,D,E} \ket{\psi_k^+, \csd}  \, ,\\
        \mathcal{T}^{2\pi}(k) & \equiv \bra{\psi_k^-, \csd} V^{(0)}_{2\pi} \ket{\psi_k^+, \csd}  \, .
    \end{aligned}
\end{equation}
we can then write $\delta_1^{(2)}$ and $\varepsilon^{(2)}$ as
\begin{equation}
    \begin{aligned}
        \delta_1^{(2)}(k) &= C^{(2)}\theta_C(k) + D^{(0)}\theta_D(k) + E^{(0)}\theta_E(k) + \theta_{2\pi}(k) \, ,\\
        \varepsilon^{(2)}(k) &= C^{(2)}\mathcal{E}_C(k) + D^{(0)}\mathcal{E}_D(k) + E^{(0)}\mathcal{E}_E(k) + \mathcal{E}_{2\pi}(k) \, , \label{eqn:NNLODeltaEpsilon}
    \end{aligned}
\end{equation}
where
\begin{equation}
    \begin{aligned}
        \theta_{C,D,E,2\pi} &\equiv \frac{-\pi k}{2\mathrm{cos}(2\varepsilon^{(0)})}Re\left[e^{-2i\delta_1^{(0)}}\mathcal{T}^{C,D,E,2\pi}_{00}(k)\right] \, , \\
        \mathcal{E}_{C,D,E,2\pi} &\equiv \frac{-\pi k}{2\mathrm{cos}(2\varepsilon^{(0)})}Re\left[e^{-i(\delta_1^{(0)}+\delta_2^{(0)})}\mathcal{T}^{C,D,E,2\pi}_{02}(k)\right] \, . 
    \end{aligned}
    \label{equ:theta_T}
\end{equation}

In addition to the phase shifts, we use the deuteron binding energy as a constraint to determine the {\NNLO} LECs. The {\NNLO} correction to the deuteron binding energy $B^{(2)}$ is calculated perturbatively using the LO deuteron wave function $\psi_L^{(0)}(p)$ obtained from Eq.~\eqref{equ:BDELO}: 
\begin{equation}
    \begin{aligned}
        B^{(2)} &\equiv \braket{\psi^{(0)}|V^{(2)}|\psi^{(0)}} \\
        &=  \frac{1}{m_N}\sum_{L, L'} \int \int dp'\, p'^2 dp\, p^2 \psi_{L'}^{(0)}(p') V^{(2)}_{L'L}(p',p)\psi_L^{(0)}(p) \, ,
    \end{aligned}
\end{equation}
with $L = L' = 0,2$. $B^{(2)}$ can also be split into four parts: 
\begin{equation}
    B^{(2)} =  C^{(2)}\beta_C + D^{(0)}\beta_D + E^{(0)}\beta_E + \beta_{2\pi} \, , \label{eqn:NNLOB}
\end{equation}
where
\begin{equation}
    \begin{aligned}
        \beta_{C,D,E} & \equiv \braket{\psi^{(0)}|\VSOI{C,D,E}|\psi^{(0)}} \, ,\\
        \beta_{2\pi} & \equiv \braket{\psi^{(0)}|V^{(0)}_{2\pi}|\psi^{(0)}} \, .
    \end{aligned}
\end{equation}

\section{\texorpdfstring{{\NNLO}}{NNLO} LECs and exceptional cutoffs\label{sec:exceptional}}

To determine the values of $C^{(2)}$, $D^{(0)}$ and $E^{(0)}$ at {\NNLO}, we construct a system of linear equations by simultaneously matching the EFT and the partial-wave analyses (PWA) from the Nijmegen group~\cite{nn.online, Stoks:1993tb} on (1) the $^3S_1$ phase shift at $k_1$ and (2) the mixing angle $\varepsilon$ at $k_2$. In addition, we also match the EFT and experimental value of the deuteron binding energy $B_\text{emp}$:
\begin{equation}
    \begin{aligned}
        \delta_1^\text{PWA}(k_1) - \delta^{(0)}_1(k_1) - \theta_{2\pi}(k_1) &= C^{(2)} \theta_C(k_1) + D^{(0)} \theta_D(k_1) + E^{(0)} \theta_E(k_1) \, ,\\
        \varepsilon^\text{PWA}(k_2) - \varepsilon^{(0)}(k_1) - \mathcal{E}_{2\pi}(k_2) &= C^{(2)} \mathcal{E}_C(k_2) + D^{(0)} \mathcal{E}_D(k_2)+ E^{(0)} \mathcal{E}_E(k_2) \, ,\\
        B_\text{emp} - B^{(0)} - \beta_{2\pi} &= C^{(2)}\beta_C + D^{(0)}\beta_D + E^{(0)}\beta_E \, .
    \end{aligned}
    \label{equ:linear_eq}
\end{equation}

Owing to the long-range attractive singular potential at LO, a careful examination of this sort of linear systems shows that the coefficient matrix can be singular for the ``exceptional cutoffs,'' as reported by the study on the uncoupled-channel cases in Ref.~\cite{Gasparyan:2022isg}.
To identify these exceptional cutoff $\Lambda$ values, we need to study the determinant of the linear system as a function of $\Lambda$:
\begin{equation}
    \mathcal{G}(\Lambda; k_1, k_2, B^{(0)})  \equiv \Lambda^{-7} \times \begin{vmatrix}
        \theta_C(k_1) & \theta_D(k_1) & \theta_E(k_1)  \\
        \mathcal{E}_C(k_2) & \mathcal{E}_D(k_2) & \mathcal{E}_E(k_2) \\
        \beta_C & \beta_D & \beta_E \\
    \end{vmatrix} \, ,
    \label{equ:G_Lambda}
\end{equation}
where the factor of $\Lambda^{-7}$ is put in place to make $\mathcal{G}$ dimensionless.
The behavior of $\mathcal{G}$ as a function of $\Lambda$ is dictated by the combination of the LO renormalization condition $B^{(0)}$ and the {\NNLO} fitting kinematic points $k_1$ and $k_2$, referred to in the paper as a ``fitting scheme.''

\begin{figure}
    \centering
    \includegraphics[width = 0.3\textwidth]{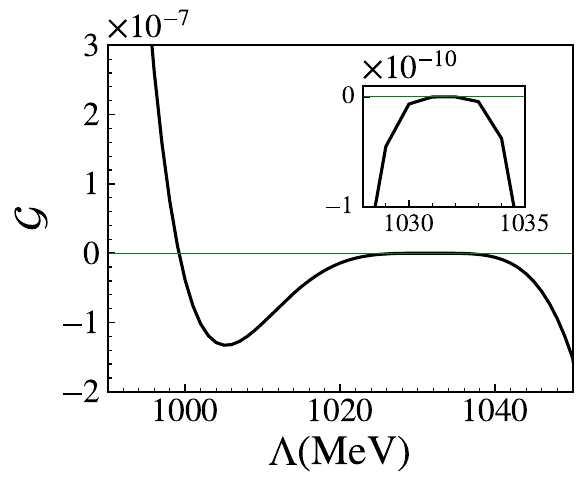}
    \includegraphics[width = 0.3\textwidth]{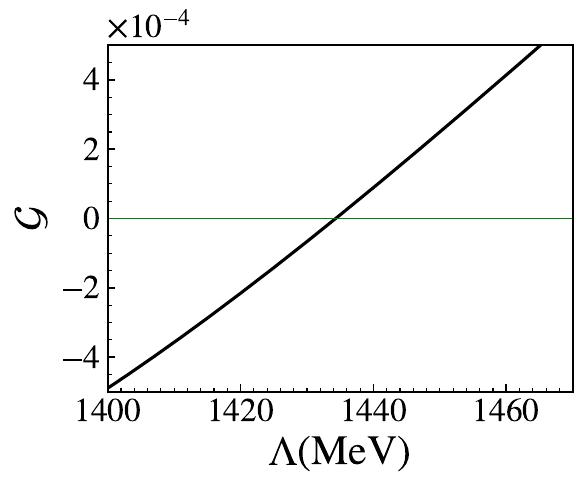}
    \includegraphics[width = 0.3\textwidth]{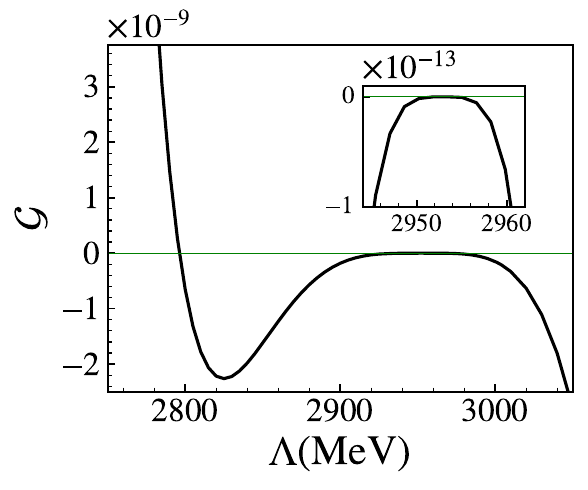}
`    \caption{$\mathcal{G}(\Lambda; k_1, k_2, B^{(0)})$ as a function of $\Lambda$ using Scheme I everywhere.}
    \label{fig:detA_300_200Original}
\end{figure}

When $\mathcal{G}(\Lambda; k_1, k_2, B^{(0)})$ vanishes at a certain cutoff it may become an issue to determine the {\NNLO} LECs.
But not all the zeros of $\mathcal{G}(\Lambda; k_1, k_2, B^{(0)})$ concern us.
At $\Lambda_\star$ where $C^{(0)}$ diverges, $\mathcal{G}$ always vanishes as $\mathcal{G}(\Lambda; k_1, k_2, B^{(0)}) \to (\Lambda - \Lambda_\star)^4$, regardless of the values of $(k_1$, $k_2)$.
But this is not problematic.
$\theta_{C,D,E}$, $\mathcal{E}_{C,D,E}$ and $\beta_{C,D,E}$ are all approaching zero as $\Lambda \to \Lambda_\star$. Although $C^{(2)}$, $D^{(0)}$ and $E^{(0)}$ are diverging, the right-hand side of Eq.~\eqref{equ:linear_eq} remains finite. 
The nonissue of $\Lambda_\star$ for the uncoupled channels is discussed in Refs.~\cite{Gasparyan:2022isg, Peng:2024aiz}.
For the coupled channels we explain in more detail in the Appendix the characteristic behaviors of $\mathcal{G}(\Lambda; k_1, k_2, B^{(0)})$ near $\Lambda_\star$ and why the phase shifts remain finite while the {\NNLO} LECs diverge.

We focus instead on the other zeros of $\mathcal{G}$, defined as the GECs in Ref.~\cite{Gasparyan:2022isg}.
To have an idea about how they distribute along the $\Lambda$ axis, we plot $\mathcal{G}(\Lambda; k_1, k_2, B^{(0)})$ as a function of $\Lambda$ in Fig.~\ref{fig:detA_300_200Original} with the following scheme:
$k_1=300.2$ MeV, $k_2=201.0$ MeV, and $B^{(0)} = B_\text{emp}$.
For the sake of our discussions, we refer to this particular choice as Scheme I.
The plots show that $\mathcal{G}(\Lambda; k_1, k_2, B^{(0)})$ has five zeros for $\Lambda \leqslant 3200$ MeV (only their vicinities are shown), distributed across three characteristic regions: two pairs near $1020$ MeV and $2900$ MeV and an isolated zero at $1434$ MeV. 
For each of the pairs, the zero at the larger $\Lambda$ where $\mathcal{G}(\Lambda)$ becomes tangent to the $\Lambda$-axis corresponds to $\Lambda_\star$ because $\mathcal{G} \to (\Lambda - \Lambda_\star)^4$.
The lower-$\Lambda$ zero and the isolated one where $\mathcal{G}(\Lambda)$ crosses the $\Lambda$-axis are GECs.
In this paper, we will study $\Lambda_E = 1000$ and $1434$ MeV as the examples.

The vanishing $\mathcal{G}(\Lambda_E; k_1, k_2, B^{(0)})$ implies that the $\cs{3}{1}$ phase shift $\delta_1$ at $k_1$, $\varepsilon$ at $k_2$, and deuteron binding energy are correlated in such a way that the third observable can be predicted once the other two are known.
This correlation acts like some sort of symmetry and is only realized at $\Lambda_E$.
$\Lambda_E$ will likely shift to a different value if a different regulator is used.
At least for the uncoupled channel of $\chp{3}{0}$, this has been confirmed to be the case~\cite{Yang:2024yqv}. 
Not only does this correlation depend on the choice of regularization, but more importantly, it is not realized by $NN$ scattering data.
In our view, the said correlation is therefore an artifact accidentally emerging from the ultraviolet regularization.
Forcing this correlation upon the EFT will result in wrong values of the LECs, and will lead to wrong predictions at other $k$'s.
This is the phenomenon reported in Ref.~\cite{Gasparyan:2022isg} and reproduced in Ref.~\cite{Yang:2024yqv}.
In Fig.~\ref{fig:pvL_1000_mixed} the dashed lines show that the ill-defined values of the {NNLO} LECs lead to divergence of the phase shifts.

\section{Solution \label{sec:solution}}

We wish to determine the values of the {\NNLO} LECs around $\Lambda_E$ by removing the unwanted, artificial correlation.
This can be achieved by somehow changing the value of $\mathcal{G}(\Lambda; k_1, k_2, B^{(0)})$.
We advocate in Ref.~\cite{Peng:2024aiz} to modify slightly the fitting scheme in the problematic cutoff window, which makes $\mathcal{G}(\Lambda; k_1, k_2, B^{(0)})$ avoid crossing the $\Lambda$ axis by forcing it to vary discontinuously. 
The justification behind the change of the fitting scheme is that at a given order in an EFT, the truncation error grants us a leeway to choose what to fit to, so long as the EFT improves order by order.
The resulting EFT amplitude exhibits discontinuous variations in the GEC windows due to the discontinuity of $\mathcal{G}(\Lambda; k_1, k_2, B^{(0)})$, but such variations are comparable or even smaller than the EFT truncation error expected at that order.

\begin{figure}
    \centering
    \includegraphics[width = 0.4\textwidth]{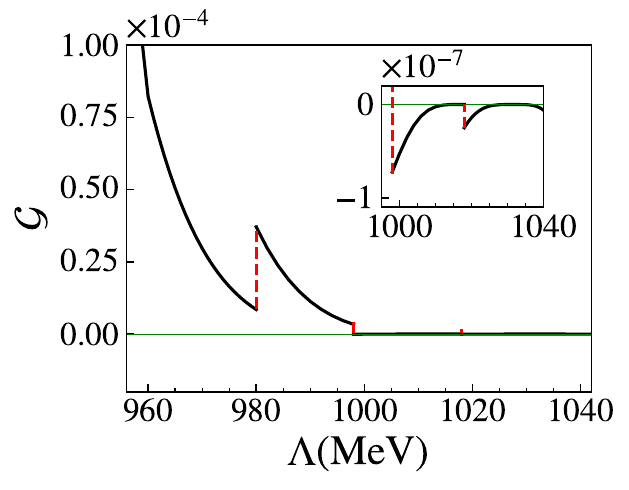}
    \includegraphics[width = 0.4\textwidth]{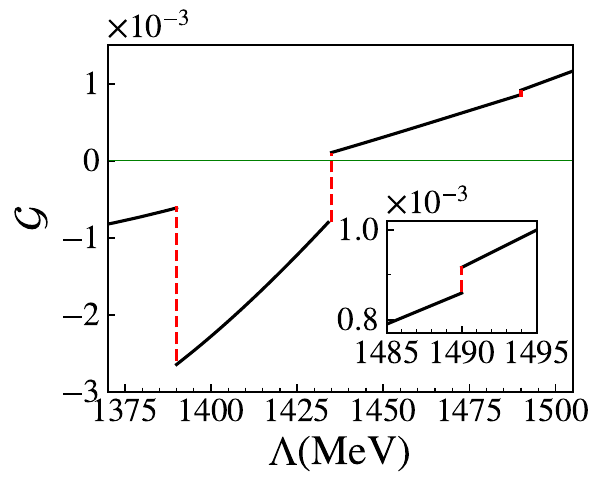}
    \caption{$\mathcal{G}(\Lambda; k_1, k_2, B^{(0)})$ as a function of $\Lambda$ using the modified fitting strategy defined in Table~\ref{tab:NNLO_k1k2}.}
    \label{fig:detA_300_200_mixed}
\end{figure}

How to modify the fitting strategy is rather an engineering problem. 
In the case of $\chp{3}{0}$~\cite{Peng_2020}, only the LO renormalization condition is modified.
The coupled channel of $\csd$ is more complicated, and we find that it is useful to modify as well $k_1$ and $k_2$ where one takes inputs from the PWA at {\NNLO}.
We state the modified fitting strategy and then explain why it works.
Table~\ref{tab:NNLO_k1k2} lists what values of $B^{(0)}$, $k_1$, and $k_2$ are used in each scheme and to which cutoff window they are applied.
The resulting $\mathcal{G}(\Lambda; k_1, k_2, B^{(0)})$ is plotted as a function of $\Lambda$ in Fig.~\ref{fig:detA_300_200_mixed}, to be compared with the original fitting strategy in which Scheme I is applied everywhere, as shown in Fig.~\ref{fig:detA_300_200Original}.
Figure~\ref{fig:detA_300_200_mixed} shows 
that both GEC zeros have been successfully removed; therefore, we can extract the well-defined LECs in the vicinity of $\Lambda_E$.
Note that in the left inset of Fig.~\ref{fig:detA_300_200_mixed} there are now two $\Lambda_\star$ located at $1018$ MeV and $1032$ MeV, identified easily by their quartic behavior $(\Lambda - \Lambda_\star)^4$.

\begin{table}[tb]
\caption{
The modified fitting strategy is composed of several schemes. Each scheme is defined by $B^{(0)}$, $k_1$, and $k_2$.
The right-most column specifies which cutoff region each scheme is applies. $\Delta B=0.5$ MeV.
}
\centering
\begin{tabular}{c|c|c|c||c}
\hline
Scheme\; &\; $B^{(0)}$(MeV) \; &\; $k_1$(MeV) \; &\; $k_2$(MeV) \; &\;$\Lambda$(MeV)\\
\hline
\hline
 I & $B_\text{emp}$ & 300.2 & 201.0 & the rest \\
\hline
 II & $B_\text{emp}+\Delta B$ & 300.2 & 201.0 & {[980,998]} \\
\hline
 III & $B_\text{emp}-\Delta B$  & 300.2 & 201.0 & {[998,1018]} \\
\hline
 IV & $B_\text{emp}$  & 350.7 & 300.2 & {[1390,1435]} \\
\hline
 V & $B_\text{emp}$  & 350.7 & 160.7 & {[1435,1490]} \\
\hline
\hline
\end{tabular}
\label{tab:NNLO_k1k2}
\end{table}

\begin{figure}
    \centering
    \includegraphics[width = 0.4\textwidth]{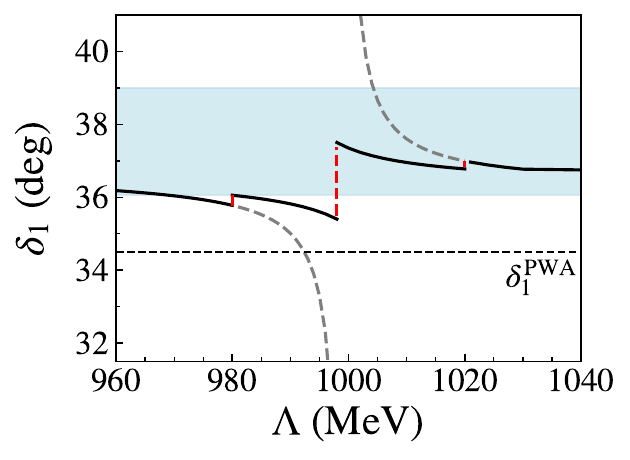}
    \includegraphics[width = 0.4\textwidth]{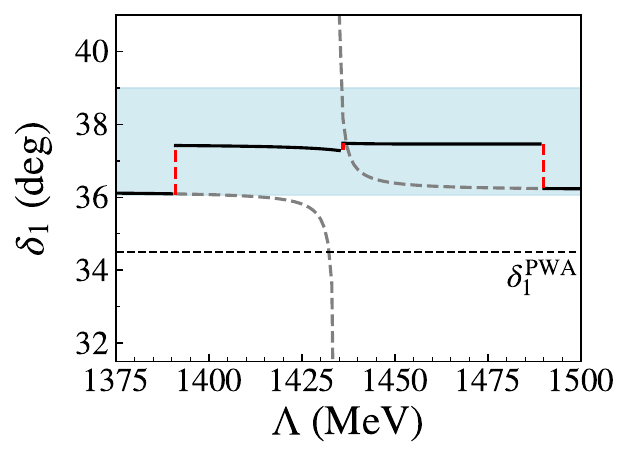}
    \caption{The {\NNLO} $\cs{3}{1}$ phase shifts at $k=250$ MeV as a function of $\Lambda$.
    The solid lines correspond to the modified fitting strategy defined in Table~\ref{tab:NNLO_k1k2}, and the dashed lines correspond to the original fitting strategy. 
    The shaded bands are the background cutoff variations. See the text for more explanations.}
    \label{fig:pvL_1000_mixed}
\end{figure}

With the modified fitting strategy, we examine the cutoff dependence of the $\cs{3}{1}$ phase shifts at {\NNLO} in Fig.~\ref{fig:pvL_1000_mixed}.
Because the phase shifts at $k_1$ and $k_2$ by construction always take the PWA values and thus are $\Lambda$ independent, we use $k = 250$ MeV which is not used in any of the fitting schemes.
For comparison, the {\NNLO} phase shifts with 
the original fitting strategy--- using Scheme I everywhere--- is represented by the dashed lines.
The divergent behavior of the dashed lines echos the pathology caused by the ill-defined values of the {\NNLO} LECs, as mentioned in Sec.~\ref{sec:exceptional}.
Adjusting the renormalization conditions according to Table~\ref{tab:NNLO_k1k2} avoids divergences, as indicated by the solid curves, but the modified fitting strategy induces discontinuities to the cutoff variation of the phase shifts.

We argue that these small breaks are within the EFT uncertainty expected at {\NNLO}.
First, we compare them with the ``background'' variation produced by the unproblematic cutoffs and indicated as the bands, that are generated by varying $\Lambda$ from 600 MeV to 3200 MeV where Scheme I is applied, according to Table~\ref{tab:NNLO_k1k2}.
The breaks induced by adjusting the fitting scheme are smaller than or comparable with the overall variations.
Second, the breaks are comparable with the EFT truncation error at {\NNLO} that is roughly estimated to be $\delta_1^\text{PWA}(k/m_\sigma)^3 \simeq 4^\circ$ where the sigma-meson mass $m_\sigma \simeq 500$ MeV~\cite{ParticleDataGroup:2024cfk} is used as the breakdown scale.
$m_\sigma$ is already an optimistic estimate of the EFT convergence radius because alternatives like using the nucleon-delta mass splitting $\Delta \simeq 290$ MeV will only increase the uncertainty estimate and thus accommodates more easily the discontinuous cutoff variations.

At the center of the modified strategy is the decomposition of the cutoffs into various sections where different schemes are applied. This decomposition and the values of $B^{(0)}$ and $(k_1, k_2)$ for each section is not a unique choice.
One can further tweak where to and how to change the fitting scheme, as long as the resulting cutoff variation is comparable with the EFT uncertainty expected at this order.

\begin{figure}
    \centering
    \includegraphics[width = 0.4\textwidth]{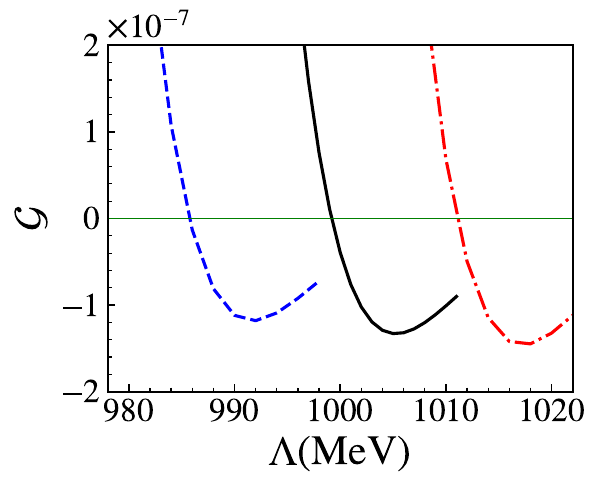}
    \includegraphics[width = 0.4\textwidth]{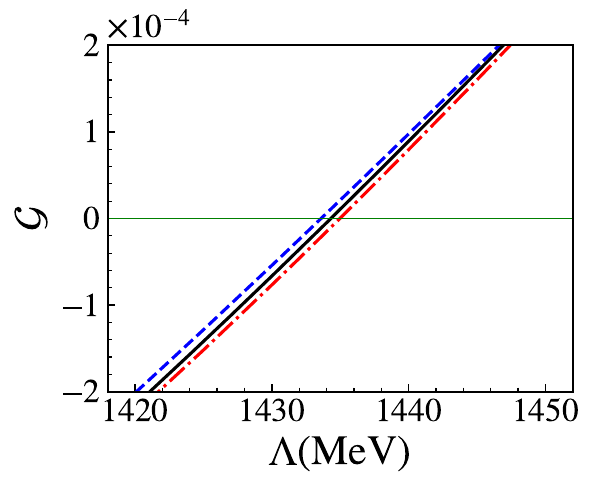}
    \caption{$\mathcal{G}(\Lambda; k_1, k_2, B^{(0)})$ versus $\Lambda$ for various values of $B^{(0)}$.
    The black solid, red dash-dotted, and blue dashed lines correspond to Scheme I, II and III, respectively.}
    \label{fig:detA_300_200_LOdiff}
\end{figure}

\begin{figure}
    \centering
    \includegraphics[width = 0.4\textwidth]{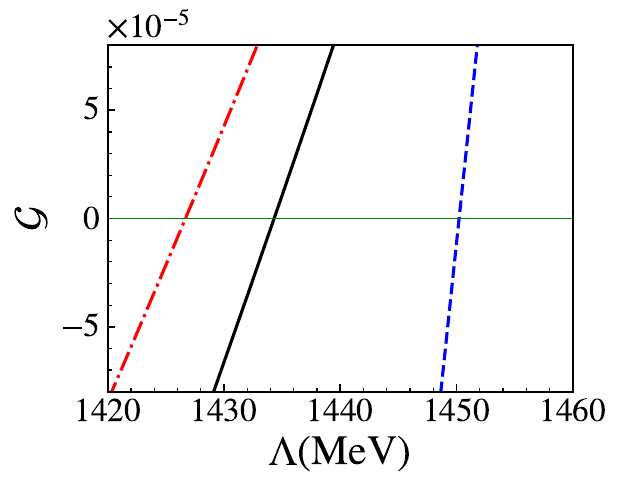}
    \caption{$\mathcal{G}(\Lambda; k_1, k_2, B^{(0)})$ versus $\Lambda$ for various values of $(k_1, k_2)$.
    The black solid, blue dashed and red dash-dotted lines correspond to Scheme I, IV and V, respectively.}
    \label{fig:detA_LOemp_NNLOdiff}
\end{figure}

In Ref.~\cite{Peng:2024aiz} where $\chp{3}{0}$ is treated, only the LO renormalization condition is adjusted, but we find it is not enough for $\csd$.
As shown in Table~\ref{tab:NNLO_k1k2}, $B^{(0)}$ is adjusted  near $1000$ MeV whereas $(k_1, k_2)$ for the {\NNLO} PWA inputs are modified near $1434$ MeV.
This is because the GEC zero of $\mathcal{G}(\Lambda)$ near $1434$ MeV is more sensitive to the variation of $(k_1, k_2)$ than to that of $B^{(0)}$.
To demonstrate this observation, we plot $\mathcal{G}(\Lambda)$ in Fig.~\ref{fig:detA_300_200_LOdiff} for three values of $B^{(0)}$: $B_\text{emp}$ and $B_\text{emp} \pm \Delta B$ where $\Delta B = 0.5$ MeV.
The $1000$-MeV zero varies significantly, by a shift of $\simeq \pm 12$ MeV, whereas the $1434$-MeV zero shifts only about $\pm1$ MeV.
Adjusting $(k_1, k_2)$ instead can make the $1434$-MeV zero move more significantly. 
As shown in Fig.~\ref{fig:detA_LOemp_NNLOdiff}, when using $B^{(0)} = B_\text{emp}$ and applying $(k_1, k_2)$ as specified in Schemes IV and V respectively, we can shift the $1434$-MeV zero considerably, by $\simeq 24$ MeV.
Although we do not fully understand the mechanism by which the renormalization conditions impact the GEC zero in $\csd$, we wish to note an important feature of the coupled-channel dynamics.
In $\csd$, the long-range singular attraction of the OPE tensor force acts as a $SD$ mixing matrix element whereas the short-range counterterm $C_0$ resides in $\cs{3}{1}$-$\cs{3}{1}$, see Eq.~\eqref{eqn:LOVS}.
Therefore, $C_0$ counteracts the OPE tensor force only through quantum fluctuations, unlike in $\chp{3}{0}$.

\begin{figure}
    \centering
    \includegraphics[width = 0.3\textwidth]{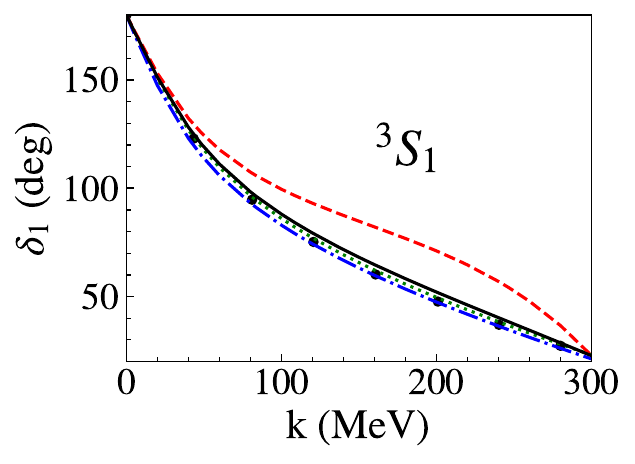}
    \includegraphics[width = 0.3\textwidth]{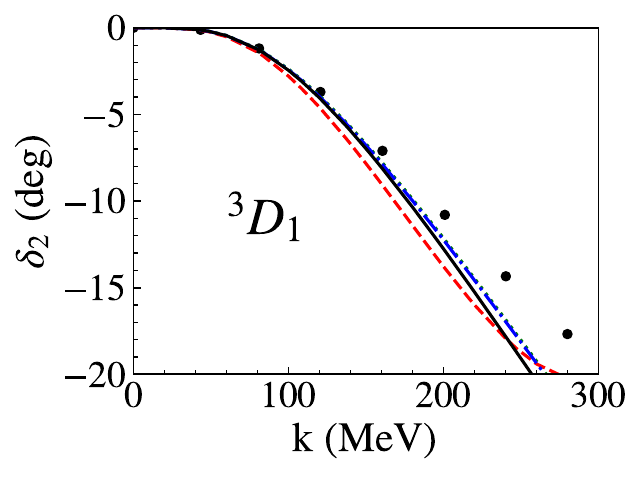}
    \includegraphics[width = 0.3\textwidth]{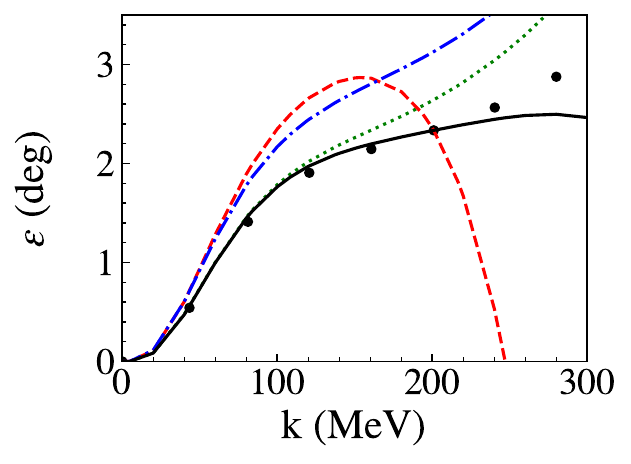}
    \caption{The $\csd$ phase shifts as functions of $k$ for $\Lambda = 1000$ MeV. The green dotted (LO) and the red dashed ({\NNLO}) curves use the original fitting strategy while the blue dash-dotted (LO) and black solid ({\NNLO}) curves use the modified one.
    The solid dots are from the PWA.}
    \label{fig:NNLO_pvk_1000}
\end{figure}

\begin{figure}
    \centering
    \includegraphics[width = 0.3\textwidth]{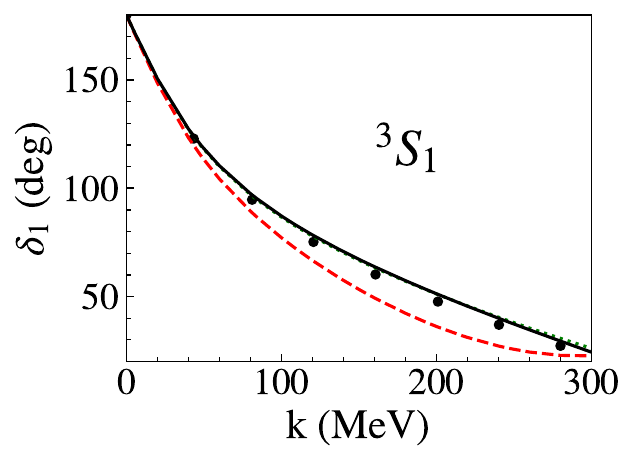}
    \includegraphics[width = 0.3\textwidth]{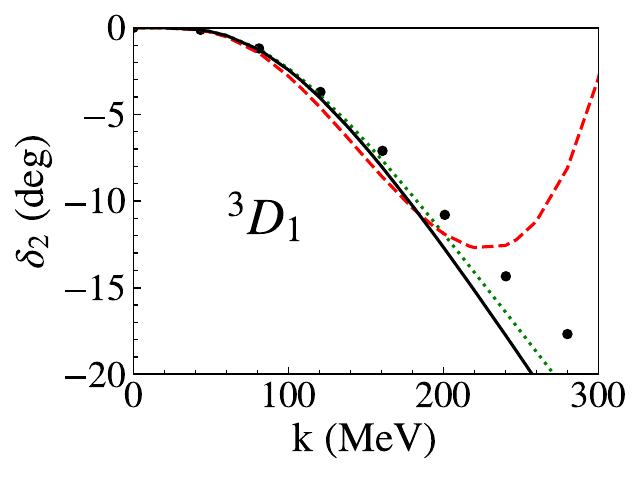}
    \includegraphics[width = 0.3\textwidth]{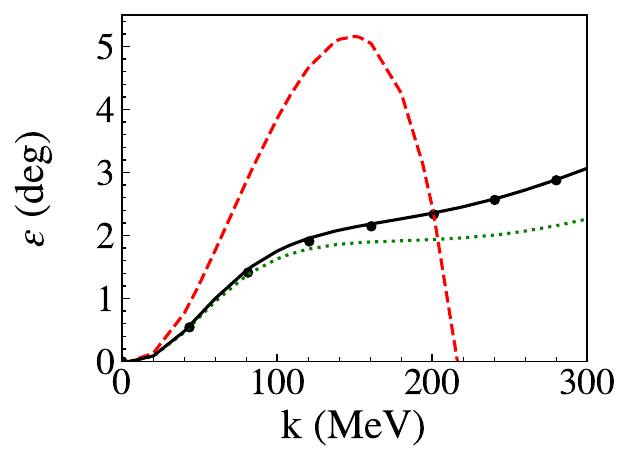}
    \caption{The $\csd$ phase shifts as functions of $k$ for $\Lambda = 1434$ MeV.
    The green dotted curves are the LO.
    At {\NNLO}, the red dashed represent the original fitting strategy and the black solid curves use the modified one.
    The solid dots are from the PWA.}
    \label{fig:NNLO_pvk_1434}
\end{figure}

We need to examine the overall comparison of the EFT phase shifts with the PWA across the chiral-EFT region.
Figures~\ref{fig:NNLO_pvk_1000} and \ref{fig:NNLO_pvk_1434} present the phase shifts calculated using the modified fitting strategy at $\Lambda_E = 1000$ MeV and $1434$ MeV, with the phase shifts from the original strategy included.
The {\NNLO} results from the modified strategy achieve good agreement with the PWA up to $k \leqslant 300$ MeV, while the original strategy shows significant deviation from the PWA.

\begin{figure}
    \centering
    \includegraphics[width = 0.3\textwidth]{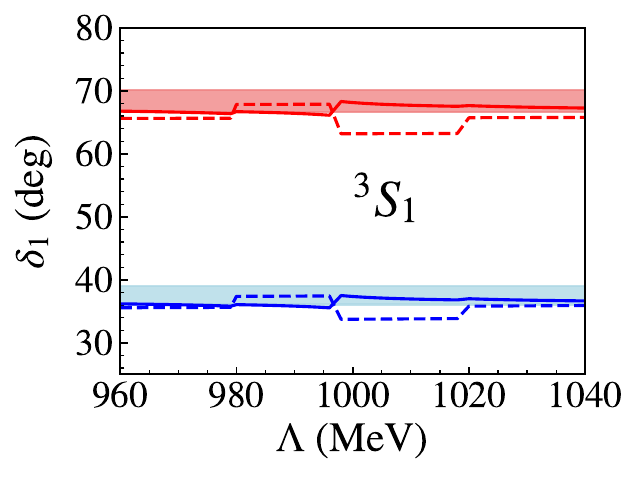}
    \includegraphics[width = 0.3\textwidth]{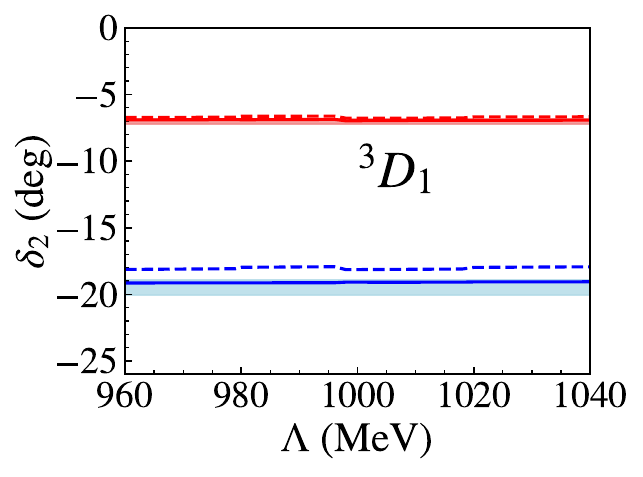}
    \includegraphics[width = 0.3\textwidth]{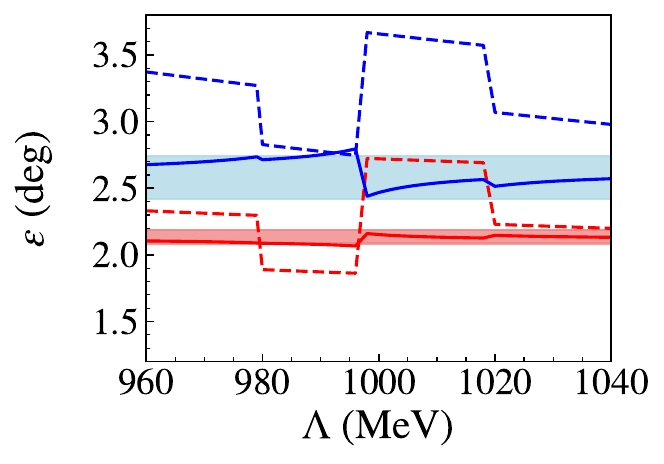}
    \includegraphics[width = 0.3\textwidth]{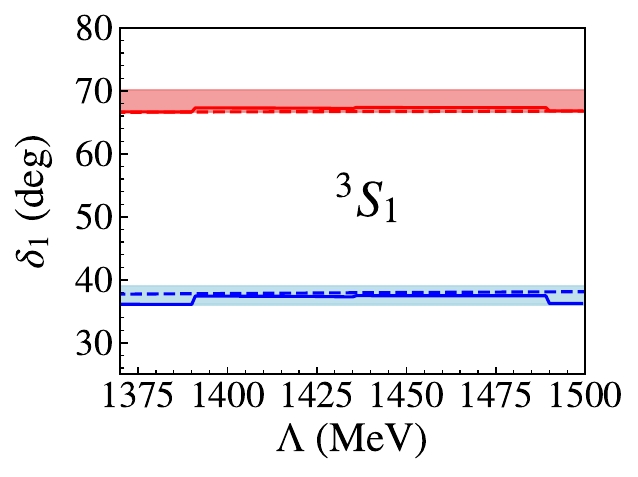}
    \includegraphics[width = 0.3\textwidth]{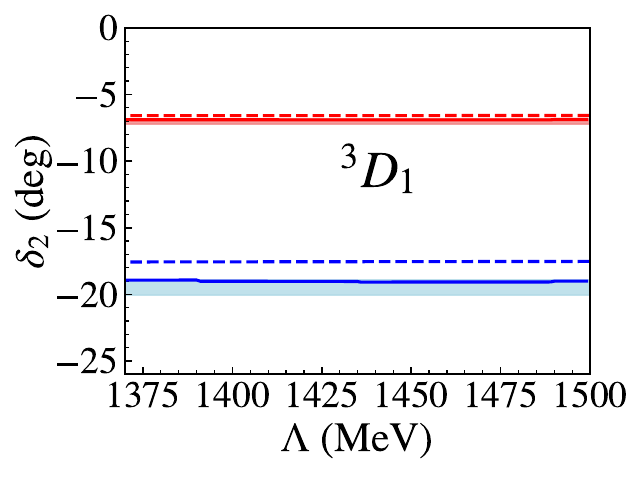}
    \includegraphics[width = 0.3\textwidth]{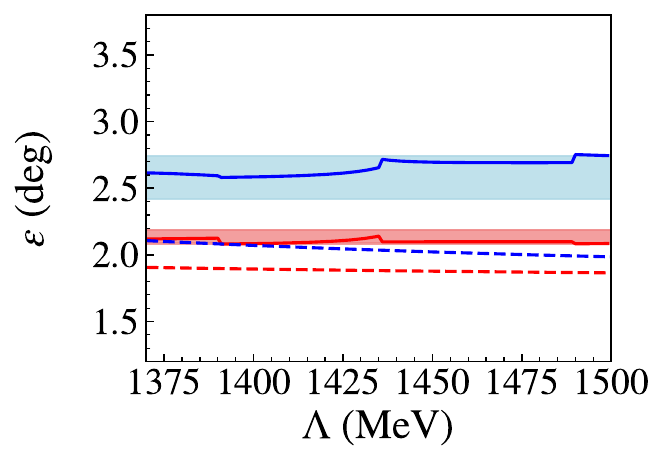}
    \caption{The $\csd$ phase shifts as functions of $\Lambda$ at a given $k$ using the modified fitting strategy. The dashed and solid curves correspond to the LO and {\NNLO} phase shifts, respectively. Different colors correspond to different values of $k$ used: blue for $k = 150$ MeV and red for $k = 250$ MeV. The bands are the background cutoff variations.}
    \label{fig:NNLO_pvL_joint}
\end{figure}

\begin{figure}
    \centering
    \includegraphics[width = 0.3\textwidth]{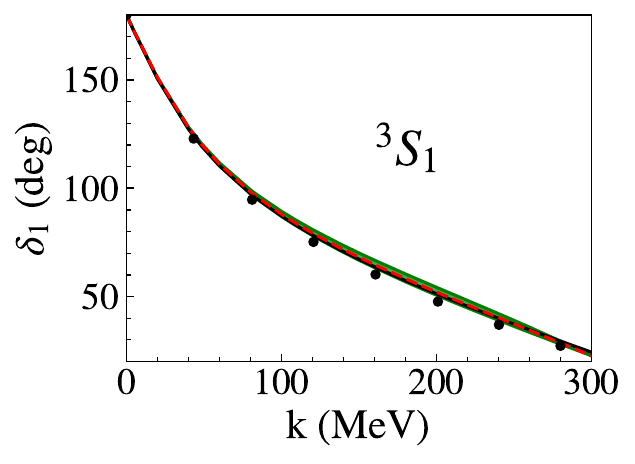}
    \includegraphics[width = 0.3\textwidth]{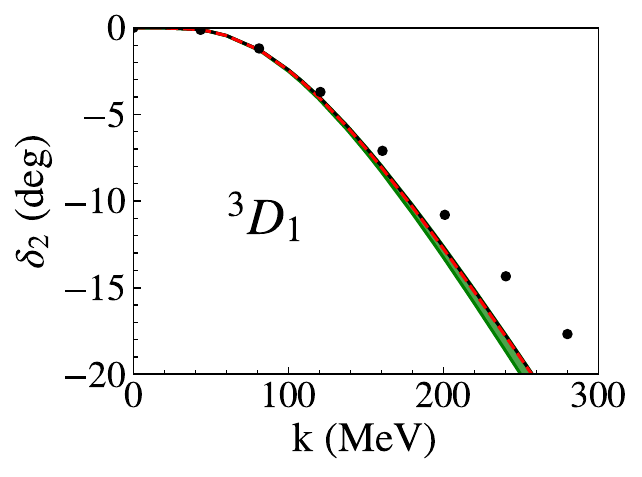}
    \includegraphics[width = 0.3\textwidth]{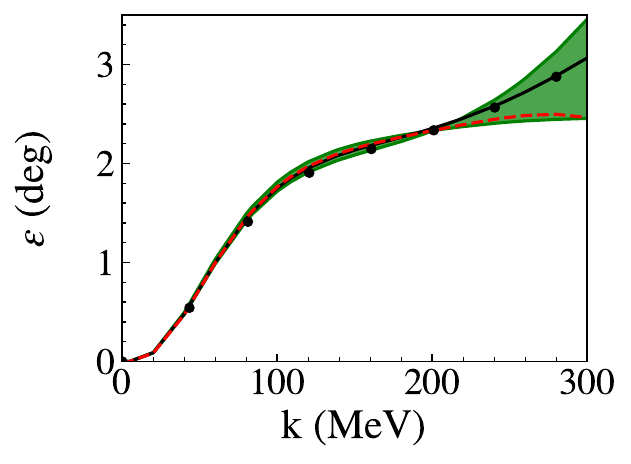}
    \caption{The {\NNLO} $\csd$ phase shifts as functions of $k$. The red dashed and the black solid lines represent the {\NNLO} using the modified fitting strategy for $\Lambda = 1000$ MeV and $\Lambda = 1434$ MeV, respectively.
    The green bands are the background variations.
    The solid dots are from the PWA.}
    \label{fig:NNLO_pvk_band}
\end{figure}

To conclude the section, we investigate more thoroughly the cutoff dependence of the EFT phase shifts.
Figure~\ref{fig:NNLO_pvL_joint} presents the LO and {\NNLO} phase shifts as functions of $\Lambda$ for $k = 150$ MeV and $250$ MeV.
The bands are again the background cutoff variations, generated in the same way as those bands in Fig.~\ref{fig:pvL_1000_mixed}.
The LO bumps near $1000$ MeV, induced by the modified $B^{(0)}$, are considerably suppressed at {\NNLO}.
Minor {\NNLO} fluctuations around 1434 MeV originate from employing different values of $(k_1, k_2)$.
What is important is that the fluctuations near the GECs are comparable with the background cutoff variation.
To verify that this is the case across the whole EFT kinematic region, we plot the background cutoff variation, as the green bands, and the {\NNLO} phase shifts as functions of $k$ for $\Lambda = 1000$ and $1434$ MeV in Fig.~\ref{fig:NNLO_pvk_band}. 
These plots indicate clearly that the EFT phase shifts curves are within the background variation bands.
We would like to note that these background variation bands are at least comparable to, if not significant smaller than, the EFT truncation error at {\NNLO}.
For instance, the background variation of the $^3S_1$ phase shift at $k=250$ MeV is about $3^\circ$, in comparison with $4^\circ$ estimated previously when we comment on Fig.~\ref{fig:pvL_1000_mixed}.

\section{summary\label{sec:sum}}

In renormalization of ChEFT for $NN$ scattering, one treats the LO long-range potential of OPE nonperturbatively and does subleading potentials in the distorted-wave expansion.
When OPE is singular and attractive in a given channel, $\csd$ in our case, the genuine exceptional cutoffs are the cutoff values where correlations between the subleading phase shifts at two kinematic points are induced.
In our view, such correlations are artifacts from the ultraviolet regulator because they are not observed in $NN$ scattering data and are sensitive to the choice of regulators and renormalization conditions.
It is necessary to remove these artificial correlations because they produce ill-defined LECs, causing the subleading EFT phase shifts to diverge near the GECs. 

We have showed that the coupled channel of $\csd$ is also plagued by the GEC problem just like $\chp{3}{0}$.
The solution is in the same spirit as the one applied in the $\chp{3}{0}$ case, based on the observation that the GECs can be moved by modifying the renormalization conditions.
These conditions, including the LO deuteron binding energy, the PWA phase shift $\delta_1$ at $k_1$ and the mixing angle at $k_2$, decide how the {\NNLO} LECs are determined.
By making slight adjustments to these conditions in the GEC windows, or changing the fitting strategy, one can avoid the said artificial correlations, thereby preventing the subleading EFT phase shifts from diverging.

The criterion for any modifications on the fitting strategy is that they introduce only small cutoff variations to the phase shifts that are comparable with the EFT uncertainty expected at this order.
By trial and error, we found a fitting strategy that satisfies this criterion, as tabulated in Table~\ref{tab:NNLO_k1k2}.
Not only this modified strategy offers mild cutoff variations, it shows a much better agreement with the PWA near the GECs than the original fitting strategy does.

\acknowledgments

This work was supported by the National Natural Science Foundation of China (NSFC) under Grant Nos. 12275185, 12335002, 12347154, 12335007, and 12035001. The High-Performance Computing Platform of Peking University is acknowledged for providing computational resources.

\appendix
\section*{APPENDIX: {\NNLO} correction in $\csd$ channel near $\Lambda_\star$}
\renewcommand{\theequation}{A.\arabic{equation}}
\setcounter{equation}{0}  

With the two-potential trick~\cite{Kaplan:1996xu, Long:2012ve, Gasparyan:2022isg}, the LO amplitude $T^{(0)}$ can be rewritten in terms of the pure OPE amplitude $T^{\pi}$:
\begin{equation}
    T_{L'L}^{\pi}(p',p;k)= V^{\pi}_{L'L}(p',p)+\sum_{L''=0,2}\int dl\, l^2V^{\pi}_{L'L''}(p',l) G_0(l;k) T_{L''L}^{\pi}(l,p;k) \, .
\end{equation}
where $G_0(l;k) \equiv (k^2-l^2+i\epsilon)^{-1}$ is the nonrelativistic propagator of free $NN$ states. 
$T^{(0)}$ can then be reconstructed as: 
\begin{equation}
    T^{(0)}_{L'L}(p',p;k)=T^{\pi}_{L'L}(p',p;k)+\frac{\chi^{\pi}_{0L'}(p')\chi^{\pi}_{0L}(p)}{(C^{(0)})^{-1}-I_{00}(k)} \, ,    
\end{equation}
 where $\chi^{\pi}_{0L}(p;k)$ and $I_{00}(k)$ are defined as follows:
\begin{align}
    \chi^{\pi}_{0L}(p) &\equiv f_R(\frac{p^2}{\Lambda^2}) + \int dl\, l^2 f_R(\frac{l^2}{\Lambda^2}) G_0(l;k) T_{0L}^{\pi}(l, p; k)  \, , \\
    I_{00}(k) &\equiv \int dp\, p^2 f_R(\frac{p^2}{\Lambda^2})G_0(p;k) \chi^{\pi}_{00}(p)  \, .
\end{align}

\begin{figure}
    \centering
    \includegraphics[width = 0.5\textwidth]{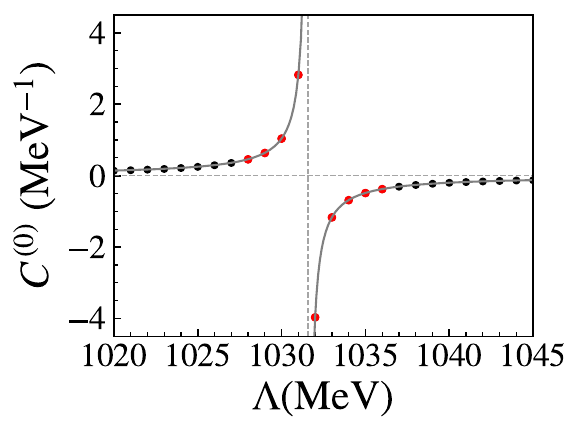}
    \caption{$C^{(0)}$ as a function of $\Lambda$.
    The solid dots are the values of $C^{(0)}$ obtained by fitting $B^{(0)}$ to $B_\text{emp}$. The solid line represents the fit using $C^{(0)} = \mu/(\Lambda-\Lambda_\star)$, and the red dots are those used in the fit.}
    \label{fig:C0_fitted}
\end{figure}
The following quantities will prove useful when the distorted-wave expansion of contact potentials is considered:
\begin{equation}
    \begin{aligned}
        \chi_{00} (k;\Lambda) &\equiv f_R(\frac{k^2}{\Lambda^2}) + \int dl\, l^2 f_R(\frac{l^2}{\Lambda^2}) G_0(l;k) T_{00}^{(0)}(l, k; k)  \,  \\
        &= \frac{\chi^\pi_{00}(k)}{1-C^{(0)}I_{00}(k)}  \, ,
    \end{aligned}
    \label{equ:chi_ll}
\end{equation}
and
\begin{equation}
    \begin{aligned}
        \phi_{0L} (k;\Lambda) &\equiv k^2f_R(\frac{k^2}{\Lambda^2}) + \int dl\, l^4 f_R(\frac{l^2}{\Lambda^2}) G_0(l;k) T_{0L}^{(0)}(l, k; k)  \,  \\
        &= \phi^\pi_{0L} (k) + \frac{\mathcal{U}_{00}(k)\chi^\pi_{0L}(k)}{1/C^{(0)}-I_{00}(k)}  \, ,
    \end{aligned}
    \label{equ:phi_0l}
\end{equation}
where 
\begin{equation}
    \begin{aligned}
        \phi^\pi_{0L} (p;k) &\equiv k^2f_R(\frac{k^2}{\Lambda^2}) + \int dl\, l^4 f_R(\frac{l^2}{\Lambda^2}) G_0(l;k) T_{0L}^\pi(l, p; k)  \,  ,\\
        \mathcal{U}_{00}(k) &\equiv \int dp\, p^2 f_R(\frac{p^2}{\Lambda^2}) G_0(p;k) \phi^{\pi}_{00}(p, k)  \, .
    \end{aligned}
\end{equation}
From nonrelativistic quantum scattering theory, one can infer that $\chi_{00}$ and $\phi_{0L}$ are proportional to the value of the wave function and its second derivative at the origin $r = 0$.

$C^{(0)}$ is shown in Ref.~\cite{Nogga:2005hy} to diverge at $\Lambda_\star$ when a spurious deep bound state appears.
The divergence of $C^{(0)}$ and vanishing of $\chi_{00}$ are indeed correlated: an additional node to the wave function suggests that a deeper state emerges.
From the right-hand side of Eq.~\eqref{equ:chi_ll}, we postulate that $\chi_{00}(k; \Lambda)$ is analytic in $\Lambda$ and thus $C^{(0)}(\Lambda)$ likely has poles at $\Lambda_\star$.
We then verify numerically that $C^{(0)} \propto (\Lambda-\Lambda_\star)^{-1}$, as shown in Fig.~\ref{fig:C0_fitted}. As a result, $\chi_{00}(k;\Lambda)$ takes the following limiting form regardless of the value of $k$:
\begin{equation}
    \begin{aligned}
        \chi_{00} (k;\Lambda) &\propto (\Lambda - \Lambda_\star)\frac{\chi^\pi_{00}(k;k)}{I_{00}(k)}\, , \label{eqn:chi00Vanish}
    \end{aligned}
\end{equation}

Using their definitions in Eq.~\eqref{equ:theta_T}, one finds that $\theta_{C,D,E}(k; \Lambda)$ factorize into products of $\chi_{00}$ and $\phi_{0L}$:
\begin{equation}
    \begin{aligned}
        \theta_{C}(k;\Lambda) &= \frac{-\pi k}{2\mathrm{cos}(2\varepsilon^{(0)})}Re\left\{e^{-2i\delta_1^{(0)}} \times [\chi_{00}(k;\Lambda)]^2\right\} \, , \\
        \theta_{D}(k;\Lambda) &= \frac{-\pi k}{2\mathrm{cos}(2\varepsilon^{(0)})}Re\left[e^{-2i\delta_1^{(0)}} \times 2\chi_{00}(k;\Lambda)\phi_{00} (k;\Lambda)\right] \, , \\
        \theta_{E}(k;\Lambda) &= \frac{-\pi k}{2\mathrm{cos}(2\varepsilon^{(0)})}Re\left[e^{-2i\delta_1^{(0)}} \times 2\chi_{00}(k;\Lambda)\phi_{02} (k;\Lambda)\right] \, ,
    \end{aligned}
    \label{equ:theta_tpt}
\end{equation}
which, when combined with Eq.~\eqref{eqn:chi00Vanish}, says that $\theta_{C,D,E}(k; \Lambda)$ vanish in certain ways near $\Lambda_\star$:
\begin{equation}
    \theta_{C}(k;\Lambda) \propto (\Lambda - \Lambda_\star)^{2} \, , \quad
    \theta_{D}(k;\Lambda) \propto (\Lambda - \Lambda_\star) \, , \quad
    \theta_{E}(k;\Lambda) \propto (\Lambda - \Lambda_\star) \, . 
\end{equation}
The same limiting behavior near $\Lambda_\star$ can be established for $\mathcal{E}_{C, D, E}$ and $\beta_{C, D, E}$.
Therefore, the {\NNLO} amplitude and binding-energy correction, defined in Eqs.~\eqref{eqn:NNLODeltaEpsilon} and \eqref{eqn:NNLOB}, can be made finite by making the {\NNLO} LECs acquire a double or simple pole at $\Lambda = \Lambda_\star$:
\begin{equation}
    C^{(2)}(\Lambda) \propto (\Lambda - \Lambda_\star)^{-2} \, , \quad
    D^{(0)}(\Lambda) \propto (\Lambda - \Lambda_\star)^{-1} \, , \quad
    E^{(0)}(\Lambda) \propto (\Lambda - \Lambda_\star)^{-1} \, . 
\end{equation}

There is another consequence of Eq.~\eqref{eqn:chi00Vanish}.
Because the matrix elements in Eq.~\eqref{equ:G_Lambda} have a simple or double zeros near $\Lambda_\star$:
\begin{equation}
    \mathcal{M}_C \propto (\Lambda - \Lambda_\star)^2, \quad  \mathcal{M}_{D,E} \propto (\Lambda - \Lambda_\star) \, ,
\end{equation}
where $\mathcal{M} = \theta$, $\mathcal{E}$, or $\beta$,
the determinant $\mathcal{G}(\Lambda)$ must have a fourth-order zero at $\Lambda_\star$:
\begin{equation}
    \mathcal{G}(\Lambda) \propto (\Lambda - \Lambda_\star)^4 \, .
\end{equation}

\bibliography{refs.bib}

\end{document}